\documentclass[10pt, conference, letterpaper]{IEEEtran}

\usepackage{graphicx}
\usepackage{siunitx}
\usepackage[dvipsnames,table,xcdraw]{xcolor}
\usepackage[hidelinks, bookmarks=false]{hyperref}
\usepackage{ulem}
\usepackage{caption}
\usepackage{amsmath}
\usepackage{tabularx}

\title{Understanding Delays in AF\_XDP-based Applications}

\author{\IEEEauthorblockN{Killian Castillon du Perron}
\IEEEauthorblockA{\textit{Université Côte d’Azur, CNRS, I3S} \\
France \\
killian.castillon-du-perron@univ-cotedazur.fr}
\and
\IEEEauthorblockN{Dino Lopez Pacheco}
\IEEEauthorblockA{\textit{Université Côte d’Azur, CNRS, I3S} \\
France \\
dino.lopez@univ-cotedazur.fr}
\and
\IEEEauthorblockN{Fabrice Huet}
\IEEEauthorblockA{\textit{Université Côte d’Azur, CNRS, I3S} \\
France \\
fabrice.huet@univ-cotedazur.fr}
}

\newcommand{\mus}[1]{\SI{#1}{\us}} 

\begin{document}

    \maketitle


    \begin{abstract}

Packet processing on Linux can be slow due to its complex network stack. To
solve this problem, there are two main solutions: eXpress Data Path (XDP) and
Data Plane Development Kit (DPDK). XDP and the AF\_XDP socket offer full
interoperability with the legacy system and is being adopted by major internet
players like Open vSwitch or Facebook. While the performance evaluation of
AF\_XDP against the legacy protocol stack in the kernel or against DPDK has
been studied in the literature, the impact of the multiple socket parameters
and the system configuration on its latency has been left aside. To address
this, we conduct an experimental study to understand the XDP/AF\_XDP ecosystem
and detect microseconds delays to better architect future latency-sensitive
applications. Since the performance of AF\_XDP depends on multiple parameters
found in different layers, finding the configuration minimizing its latency is
a challenging task. We rely on a classification algorithm to group the
performance results, allowing us to easily identify parameters with the biggest
impact on performance at different loads. Last, but not least, we show that
some configurations can significantly decrease the benefits of AF\_XDP, leading
to undesirable behaviors, while other configurations are able to reduce such
round trip delays to an impressive value of \mus{6.5} in the best case,
including the tracing overhead. In summary, AF\_XDP is a promising solution,
and careful selection of both application and socket parameters can
significantly improve performance.

    \end{abstract}
    
    
    \section{Introduction}
\label{sec:intro}

Nowadays, cloud applications are  deployed as chains of functions that need to
communicate between each other to provide full services
\cite{Gan19-deathstarbench}. While the replacement of a monolithic architecture
for the adoption of a microservice one (where several functions interconnect)
eases development and deployment, it might also bring several challenges that
need to be addressed in order to preserve the Quality of Experience required by
the customers.

Hence, latency-sensitive applications, which are increasingly popular, such as
live streaming, live gaming, and data query just to mention a few
\cite{Li22-livenet, Gan19-deathstarbench}, call for optimizations at several
layers of the service architecture. The final objective being to achieve fast
communication between functions to satisfy the customers' Quality of Experience
requirements. For instance, one single search at Google leads to thousands of
RPCs (Remote Procedure Calls)\cite{barroso17}.

At the network layer, microservices can be smartly placed to bring optimized
network overlays where communicating functions are close to each other on the
Data Center fabric. Thus, network processing and latency is reduced while
avoiding bottlenecks \cite{Li22-livenet}.

To increase the benefits from this smart microservice placement, one can deploy
fast network packet processing solutions on the servers. Note that overlay
networks are frequently composed of both hardware and software networking
devices. While the former already provides the fastest packet processing,
servers where software networking devices live offer by default only slow
packet processing capabilities \cite{zhangDemikernelDatapathOS2021,
zhuoSlimOSKernel2019}.

Two main different solutions can be adopted to speed up the processing of
network packets: DPDK  (Data Plane Development Kit) \cite{DPDKWebsite} and XDP
(eXpress Data Path) \cite{XDPProjectGitHub}. Both solutions offer fast packet
processing capabilities by sending, as fast as possible, network packets to the
user space of a system, where applications are expected to live, and where
parallelism and complex operations are easy to implement. Indeed, the packet
processing in the network protocol stack implementation of the kernel system
might be very slow due to the complexity of the current code
\cite{zhuoSlimOSKernel2019, zhangDemikernelDatapathOS2021}. The slow packet
processing can moreover be exacerbated when overlay technologies are deployed,
such as VxLAN \cite{vxlanRFC} as packets can be processed multiple times by the
entire network stack implementation~\cite{zhuoSlimOSKernel2019}.

DPDK provides a set of libraries to  create a direct path between the network
interface card (NIC) and the user space programs. By doing so, DPDK completely
\textit{steals} the packet from the kernel, with the side effect of standard
Linux networking tools not working anymore~\cite{tuRevisitingOpenVSwitch2021}.

XDP, with the help of the associated AF\_XDP socket, can be seen as an
intermediary solution between DPDK and the default network stack implementation
of current systems. AF\_XDP relies on the XDP hook which is implemented at the
NIC driver level in Linux systems to create a short path between the NIC and
user space (see Figure \ref{fig:xdpstack}). A detailed view of the AF\_XDP
architecture is provided in \S\ref{sec:xdp-afxdp-arch}. Since AF\_XDP has full
kernel support, usual networking tools (\texttt{ip}, \texttt{ping}, ...)
employed in Linux systems are fully compatible with AF\_XDP.


Understanding every factor potentially impacting the packet processing delay is
important to chase out microsecond delays in microservice-based
architectures\cite{barroso17}. We present in this paper an in depth study about
the cost of packet processing by AF\_XDP in a testbed described in
\S\ref{sec:testbed}, using two different network drivers,  Mellanox mlx5 and
Intel i40e. To do this, we generated close to 400 different configurations from
which we analysed the results using a classification algorithm to determine
what parameters lead to the best and the worst latencies (\S\ref{sec:res}). All
these aspects have not been or only partially covered by the current state of
the art (\S\ref{sec:sota}).

Our results, available in \S\ref{sec:res}, show that:
\begin{itemize}
	\item energy saving mechanism have little to no impact on the latency with 
	AF\_XDP socket with the correct parameters.
	\item with the best combination of parameters, the round-trip latency 
between two servers can reach \mus{6.5}, 
which includes an approximate 5-\mus{10} overhead due to our 
performance tracing technique.
\end{itemize}

\section{Architecture of XDP and AF\_XDP}
\label{sec:xdp-afxdp-arch}


\subsection{XDP}
\label{sec:xdp}

\textit{eBPF} (\textit{extended Berkeley Packet Filter}) is a solution to extend the capabilities of the Linux kernel, without the need to modify or write new kernel modules.
eBPF allows running sandboxed program inside hooks points located at several places of the kernel architecture.
One of those hooks is called XDP (\textit{eXpress Data Path}) and is used to 
install an eBPF program at the earliest stage of the  networking stack, inside the 
network device driver

By leveraging the XDP capabilities, firewalls and load-balancers providing very fast packet processing have been deployed on servers~\cite{FacebookincubatorKatran2023}.

\begin{figure}[htbp]
    \centerline{\includegraphics[width=0.6\columnwidth]{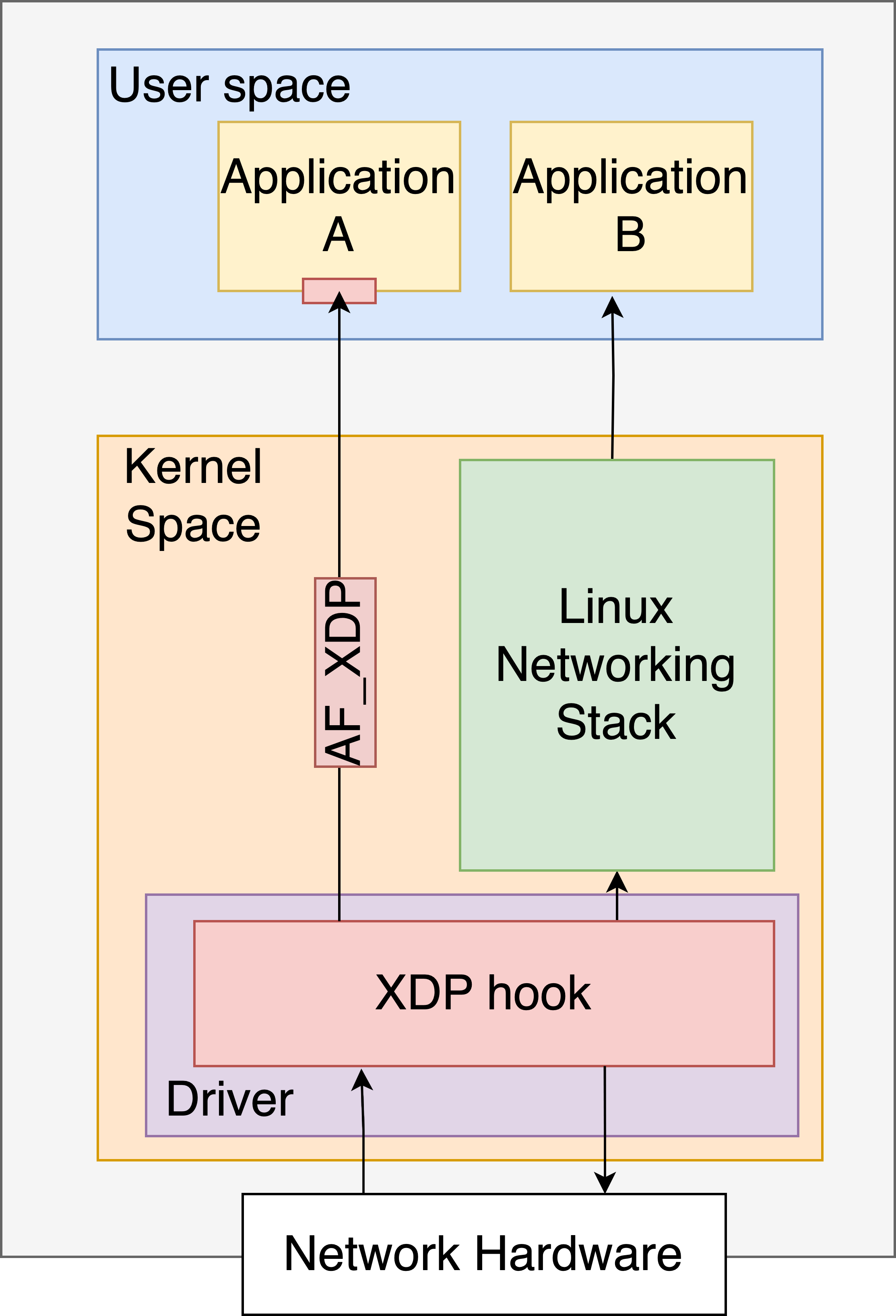}}
    \caption{XDP and AF\_XDP in the Linux networking stack.}
    \label{fig:xdpstack}
\end{figure}

\subsection{AF\_XDP}
\label{sec:af_xdp}

AF\_XDP~\cite{karlssonPathDPDKSpeeds,AFXDPLinux} is a new socket address
family which leverages XDP to receive and send packets directly between a queue
of the NIC and some user space process. It was proposed as a way to easily
bypass the legacy network protocol stack implementation, as shown in 
Figure~\ref{fig:xdpstack}. 


A memory area, called a UMEM,  is shared by both the application and the
driver. It is divided into equally sized frames which can be owned by only one
of them at a time. 

The UMEM relies on  two memory \textit{rings}, \textit{Fill} (FILL) and
\textit{Completion} (COMP), which are single consumer / single producer. They are used
to transfer ownership of the UMEM frames between user space and kernel space. The socket 
itself has two more rings, \textit{RX} and  \textit{TX}, used 
for the actual exchange of the packets.  The rings do not contain the actual packets but the addresses of the frames, as well as the length of the data for the last two rings.

Finally, the socket has two more rings, \textit{RX} and  \textit{TX}, used 
for the actual exchange of the packets.

As the Linux networking stack is bypassed, most of the processing is done either
in user-space by the application or in the driver of the NIC.
Therefore, in a simple application, most of the possible performance 
improvement will be obtained by tweaking NIC hardware and driver settings as 
well as AF\_XDP socket parameters.

    \section{Optimization parameters}

As the Linux networking stack is out of the way when using AF\_XDP, the surface
area for optimizations is further reduced to hardware and driver settings and 
the parameters of the AF\_XDP socket itself.

\subsection{Hardware and driver settings}

Many of the optimizations brought to the networking hardware and drivers are set with the goal to achieve higher rates, in terms of bits per second (bps) or packets per second (pps).
However, our goal is to identify the different components impacting latency and
its stability (jitter) independently of the throughput of the application. We
found that when optimizing for throughput,  one may induce higher latency 
especially at lower packet rates. 

\subsubsection{Interrupt delaying}

\label{sec:coalescing}
The NIC historically used to send an interrupt to the kernel for each received packet, but this 
approach doesn't work well with high packet rates~\cite{NAPIPaper}.
There are two solutions: interrupt coalescing by NAPI or by the NIC itself. NAPI (\textit{New API}) is a subsystem of the Linux Kernel that improves high-speed networking by mitigating interrupts~\cite{NAPI}.
Instead of sending an interrupt for each packet, NAPI triggers one interrupt to put the driver in \textit{poll} mode. In this mode, the driver can be later polled by the kernel to process the packets received since the last interrupt, saving CPU time. On the hardware side, \textit{coalescing} delays triggering an interrupt when receiving a packet by a certain amount of time or frames, reducing interrupt overhead. Both mechanisms involve waiting before processing a packet, which may have negligible overhead at high packet rates but can be significant at lower rates~\cite{perfsCoalescing}.

\subsubsection{Energy saving mechanisms}\label{sec:cstates}

Modern CPUs come with energy saving mechanisms, notably \textit{C-States}, which 
allow a CPU to enter a lower power state and disable some parts of itself when
idle. This can lead to higher latency and jitter as the CPU may be in a energy saving mode when receiving a packet, especially at lower rates.


\subsection{AF\_XDP socket related parameters}

The socket itself and the associated UMEM are also configurable. For the UMEM 
and the rings, the number of frames as well as their size can be configured.
However, more options are available for the socket:

\subsubsection{Zero-Copy}

By default, sending data from user space involves copying at least once the
packet between a memory region owned by the application and one owned by the
kernel for it to format the data correctly before sending it on the network.
However, network drivers supporting XDP can also add support for sending and receiving packets without
copying intermediate buffers into the NIC.  
Furthermore, zero-copy in AF\_XDP works both on the receiving and sending sides allowing to
save latency both ways~\cite{AFXDPZC}.

\subsubsection{XDP\_USE\_NEED\_WAKEUP}\label{sec:needwu}

By setting the \texttt{XDP\_USE\_NEED\_WAKEUP} option, the application enables the \texttt{need\_wakeup} flag on the TX and FILL ring. This flag is set by the driver and when false, enables the application to write and read from the rings directly without sending a syscall to the driver beforehand. When the flag is true, the application runs in the usual mode.

\subsubsection{Busy Polling}\label{sec:bp}

Typically, to signal an application that packets are available, the
underlying network driver will send IRQs (interrupts). However, this uses CPU
and can downgrade application performance if it runs on the same core as the one sending the IRQs due to context switches.
With busy polling, the application has to wake up the driver so it can start processing packets 
both on the rx and tx sides. This enables  all  processing to be performed on one core and therefore 
prevent core-to-core cache transfers which may slow down performances~\cite{afxdpBPRFC}.

\subsection{Application parameters}

\subsubsection{Polling}

What we call \textit{polling} is the use of the system call \texttt{poll()} to check for any IO activity. When we enable this option on the application, it will call \texttt{poll} regularly to process IO events.

    \section{Network Testbed}
\label{sec:testbed}

Our testbed consists of three Dell PowerEdge R520 servers with two Intel Xeon  
E5-2420 v2 CPUs at \SI{2.20}{GHz}, and 32 GB of \SI{1600}{MHz} DDR3.
Our servers feature the following NIC models:

\begin{itemize}
	\item Mellanox MT2892 ConnectX-6 \SI{100}{Gbps} with 2 ports
	\item Intel X710 \SI{10}{Gbps} with 4 ports
\end{itemize}

Those NIC are shared among the three servers (A,B,C), A with Mellanox only, C
Intel only and B having both.

All servers  are running Ubuntu 23.04 on kernel 6.2, the driver for the
Mellanox NICs is mlx5 and i40e for the Intel ones. Mellanox OFED version 23.07
has been set up on relevant servers.

For our benchmarks, servers A and C will behave as traffic generators and
receivers, while server B will only behave as a forwarder. 

\begin{figure}[htbp]
    \centerline{\includegraphics[width=\columnwidth]{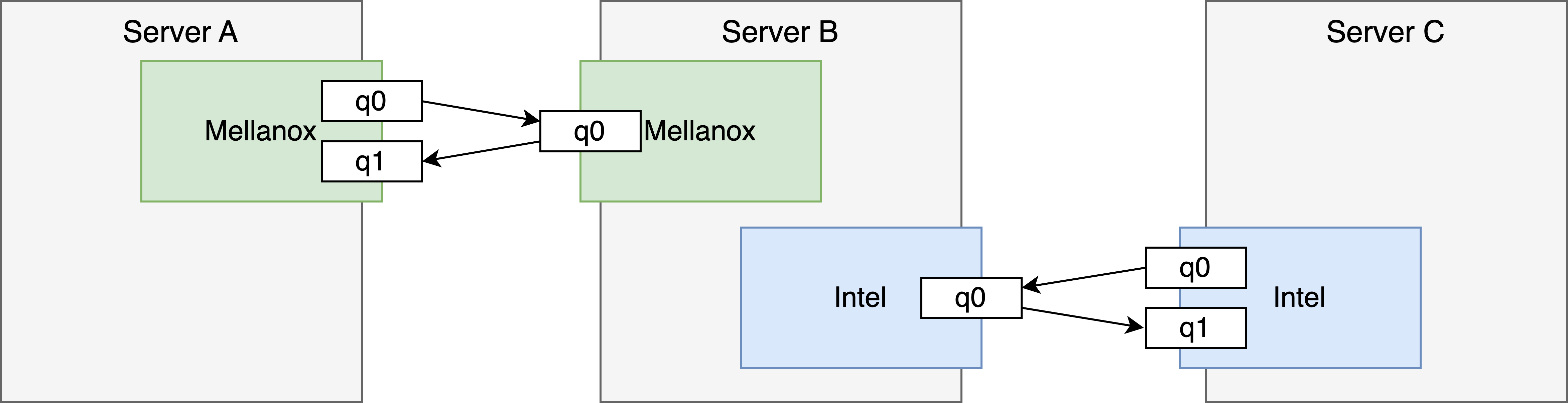}}
    \caption{Network testbed. Server A and C are traffic generators and
    receivers. Server B forwards only traffic back to the traffic generator. q0
    and q1 refer to queues.}
    \label{fig:testbed}
\end{figure}

The servers interconnection is shown in Fig.~\ref{fig:testbed}. Servers are
connected by a DAC cable, the queues (q0 and q1) being on the same port (ie.
NIC interface).

We estimated the latency difference due to the speed difference in both NIC
models to be in the order of nanoseconds in our setup, which is negligible
compared to our results. To start off, using the system's ping utility we had
\mus{191} between Mellanox hosts and \mus{211} between Intel ones.

\section{Methodology}
\label{sec:methodo}

To measure latency, we have deployed the \texttt{xdpsock}
program\cite{xdpsockGithub} on every server which is the default application
provided in Linux for testing AF\_XDP to generate, receive or forward network
packets (called \textit{pings} in this article). To gather delay-related
statistics of our network, we added USDT (Userland Statically Defined Tracing)
probes~\cite{UsingUserspaceTracepoints} to \texttt{xdpsock}. Using
\texttt{perf}~\cite{PerfWiki}, we obtain the time when batch of packets is sent
and the time of its reply. As sending and receiving is done on the same server,
we avoid clock synchronization issues when gathering timestamps.

\subsubsection{Ping parameters}

\texttt{xdpsock} generates batches of packets at fixed intervals. In our
experiments we chose to use batches of 1 packet of 64 bytes length, with a 1
second inter-batch interval. Those parameters are known to be the most
challenging at higher
rates~\cite{tuRevisitingOpenVSwitch2021,hoiland-jorgensenEXpressDataPath2018,rizzoNetmapNovelFramework2012},
which will be addressed later in our experiments.

\subsubsection{Tracing overhead}

We chose \texttt{perf} and USDT probes as they enable us to add our own probes
to \texttt{xdpsock}, while having a limited impact on the performance. Event
tracing is only performed on the sender / receiver. To estimate our tracing
overhead, we ran experiments with and without activating tracing on the
forwarder and computed the time difference between the two versions. Depending
on the inter-batch interval, the typical tracing overhead was found to be
between \mus{5} and \mus{10}.


\subsubsection{Hosts configuration and run execution}

We used Ansible~\cite{Ansible} to configure the servers before each run as well
as to run the applications and retrieve the results. This allowed us to run all
the configurations from a definition file and ensure that the system would be
set up using the correct parameters before each run:
\begin{itemize}
	\item correct number of queues set on the NIC
	\item flow steering on the NIC to the queue on which the application was listening
	\item rx and tx coalescing on the NIC
	\item selected busy polling values set in the system files
	\item setting the correct C-States level
\end{itemize}

Each benchmark is run alone, during 30 seconds and at least 3 times to ensure
consistency between runs. When running with busy polling enabled, we force the
application to run on the same core as the driver to benefit from the better
cache locality.

\section{Results}
\label{sec:res}
In this section we investigate the impact of the various optimization
parameters on the latency, for both Intel and  Mellanox cards.

To do so, we generated benchmark configurations for every combination of the
following parameters: zero-copy, busy poll, need wakeup, C-States level, rx and
tx coalescing. This produced around 100 different configurations.

In our first set of experiments, and to avoid any interference from any kind of
bottleneck, with decided to send traffic with a 1 second inter-batch interval
and batch sizes of 1 packet of 64 bytes as mentioned in \S\ref{sec:methodo}. 

\subsubsection{Measurements analysis}

During our experiments, we noticed that some configurations with zero-copy and
no polling mechanism led to very undesirable behaviors on the Intel driver,
with latencies similar to the inter-batch intervals. Therefore, we decided to
consider those configurations as outliers in our future analysis, while taking
them into considerations for our conclusion. To further clean up our traces, we
decided also to not consider configurations leading to latencies higher than
the one observed with system's pings utility which was around \mus{200} in our
testbed.

With the remaining configurations, 69 on Mellanox and 40 on Intel, we computed
general latency statistics (e.g. mean, median, standard deviations, quartiles,
minimum and maximum) for every result and averaged them out per configuration
for each NIC vendor (Intel and Mellanox).
When analyzing the results, we found that some configurations lead to similar 
latency, despite having different parameters. 

This suggested that one or more parameters coupled together lead to those particular 
results. However, identifying those clusters by hand is still challenging. 
Thus we decided to use rely on the k-means clustering algorithm to classify the 
configuration into multiple groups. 

At first, we feed the algorithm with all the metrics that we had computed: mean, 
median, standard deviation, first and third quartiles,  minimum and  maximum. 
However, the resulting clustering failed to classify the traces under good 
performances and bad performance in a clear manner.

\begin{figure}[htp]
	\centerline{\includegraphics[width=\columnwidth]{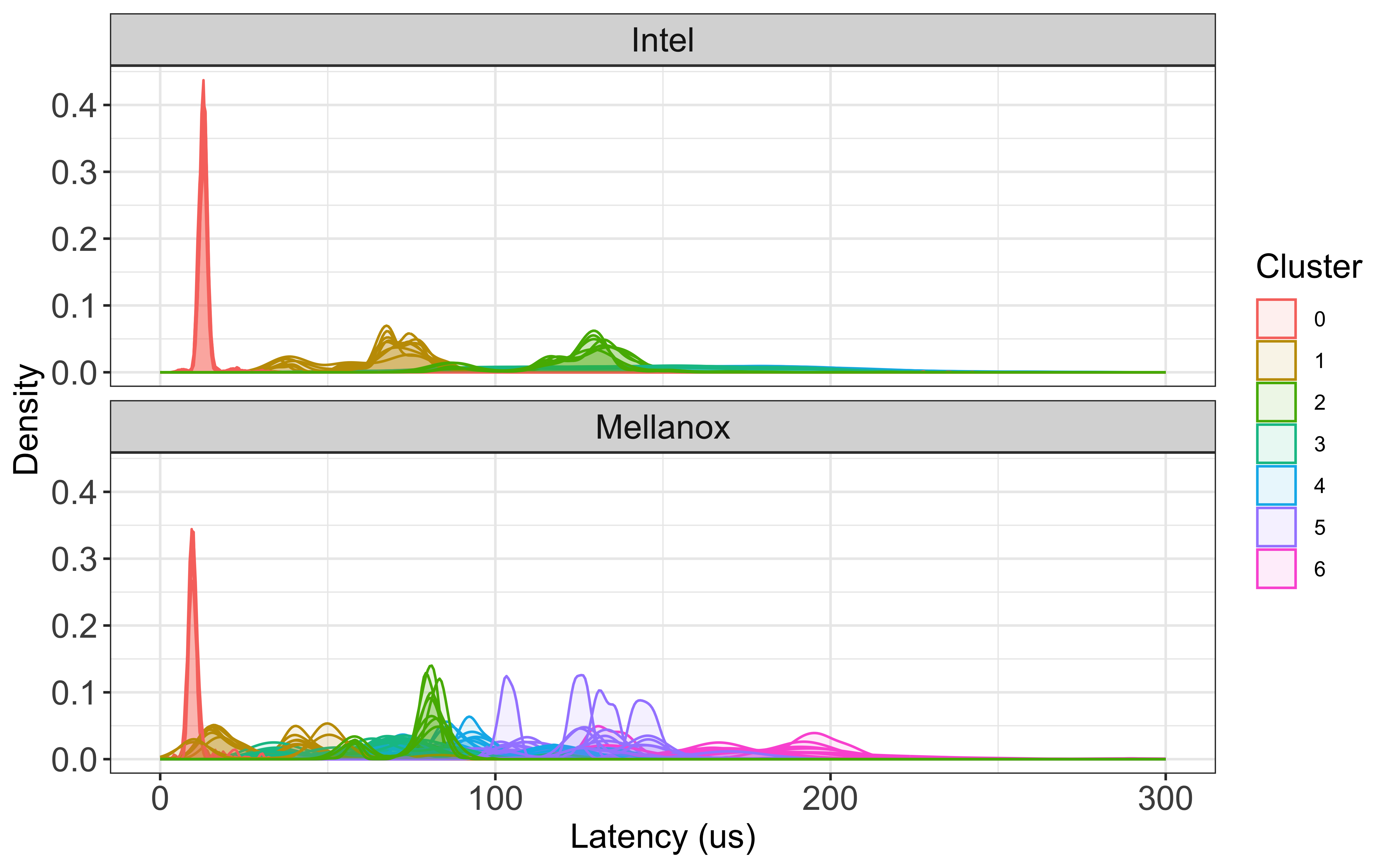}}
    \caption{Kernel density estimate plot of the top 50 configurations for both
    vendors after k-means clustering. Clusters are ordered from the lowest mean
    latency to highest.}
	\label{fig:kde-clusters}
\end{figure}

However, feeding K-means with all statistics, except the minimum and maximum
observed latencies successfully grab the performance profile. To find the
number of clusters, we wanted to obtain a good clustering based on the
following criteria: low standard deviation among clusters, small numbers of
clusters and significant statistical differences between them. This gave us a
number of 4 clusters on Intel and 6 on Mellanox. To visualize the latency
distribution of each configuration we decided to use Kernel Density Estimate
(KDE) plots, which uses a continuous probability density curve and can be
compared to a histogram but easier to read~\cite{KDE}.  As we can see in Fig.
\ref{fig:kde-clusters} where we plot the KDE of our traces assigning them to
its respective cluster, it is easy to identify each cluster on the x-axis,
which represents the average latency for a specific configuration.

According to our cluster analysis, C-States levels as well as the NIC
coalescing did not have a significant impact on the results. However, enabling
polling on the receiver always lead to worse results for both drivers.

Tables \ref{table:clusters-mellanox} and \ref{table:clusters-intel} summarize
the obtained clusters, ordered from the best configuration (cluster 0) to the
worst (cluster 5), along with the observed mean latency, standard deviation and
relevant (i.e. always present) parameters.

\begin{table}[htp!]
	\begin{center}
		\vspace{0.1in}
	\begin{tabular}{|
			>{\columncolor[HTML]{C0C0C0}}l |l|l|p{4cm}|}
		\hline
		\cellcolor[HTML]{9B9B9B}Cluster &
		\cellcolor[HTML]{9B9B9B}Mean &
		\cellcolor[HTML]{9B9B9B}Std &
		\cellcolor[HTML]{9B9B9B}Relevant parameters \\ \hline
		0 &
		\mus{10.06} & \mus{1.85} & Busy polling enabled, polling and need wakeup disabled \\ \hline
		1 & \mus{30.51}   & \mus{5.16}  & Need wakeup disabled and either one of busy polling or polling on the forwarder enabled                         \\ \hline
		2 & \mus{64.57}  & \mus{10.09} & Busy polling and need wakeup enabled\\ \hline
		3 & \mus{76.93}  & \mus{3.18}  & Polling, busy polling and need wakeup disabled\\ \hline
		4 &
		\mus{89.69} &		\mus{7.56} &
		Polling on the forwarder enabled, busy polling disabled, and either one of need wakeup and polling on the receiver enabled \\ \hline
		5 & \mus{127.41} & \mus{6.66}  & Busy polling disabled, either one of need wakeup and polling on the receiver enabled    \\ \hline
		6 & \mus{167.25} & \mus{17.274}  & Busy polling and polling on the receiver enabled    \\ \hline		
	\end{tabular}
	\caption{Clusters, statistics and relevant parameters for Mellanox at 1 packet per second. A parameter not mentioned means that it can be either enabled or disabled.}\label{table:clusters-mellanox}
	\end{center}
\end{table}

\begin{table}[htp]
	\begin{center}
		\vspace{0.1in}
	\begin{tabular}{|
			>{\columncolor[HTML]{C0C0C0}}l |l|l|p{3.8cm}|}
		\hline
		\cellcolor[HTML]{9B9B9B}Cluster & \cellcolor[HTML]{9B9B9B}Mean & \cellcolor[HTML]{9B9B9B}Std & \cellcolor[HTML]{9B9B9B}Relevant parameters \\ \hline
		0 & \mus{13.36}   & \mus{1.90}  & Busy polling enabled,  need wakeup set to false                   \\ \hline
		1 & \mus{70.00}   & \mus{6.17}  & Busy polling and need wakeup enabled                              \\ \hline
		2 & \mus{124.57}  & \mus{8.29}  & Busy polling and zero-copy disabled                               \\ \hline
		3 & \mus{139.86}  & \mus{36.91} & Need wakeup enabled, Polling enabled on the forwarder and busy polling disabled        \\ \hline
		4 & \mus{153.45}  & \mus{38.09} & Polling enabled on both forwarder and receiver, busy polling and need wakeup disabled \\ \hline
	\end{tabular}
	\caption{Clusters, statistics and relevant parameters for Intel at 1 packet per second. A parameter not mentioned means that it can be either enabled or disabled.}\label{table:clusters-intel}
	\end{center}
\end{table}


\subsubsection{Impact of load on best and worse configuration}

Finally, we wanted to test if the best and worst configurations would be
impacted by the packet rate, especially when close to the maximum line rate.

We then selected 2 configurations per cluster and re-run the tests with an
inter-packet interval close to the line rate. It was computed from minimum
latency previously obtained per vendor with equation (\ref{eq:pps}), where B is
the NIC bandwidth in bits/s, $l_{min}$ the minimal latency and P is the size of
a packet in bytes, which we set to 64. 

\begin{equation}
	\text{Packets per second} = \frac{B\times l_{min}}{8\times P}
	\label{eq:pps}
\end{equation}

This gave us a result of around 200 pps for Intel and 1750 pps for Mellanox,
which we rounded up to an inter-batch interval of \SI{5}{ms} for Intel and
\SI{0.6}{ms} for Mellanox. This led to lower latencies across all clusters,
down to a mean latency of \mus{6.5} on Mellanox and \mus{9.7} on Intel with the
best cluster, and the different identified clusters stayed relevant.

We also tested running the application without an inter-batch interval. To do
so we added a subsampling argument to our tracing function to not overload
\texttt{perf} which could alter our application's performance. This led to
higher latencies for Mellanox for the previously best results as we started to
have some buffering. On Intel however, this crashed the sender's network driver
at least for the NIC on which the program was running, preventing us from
obtaining any result at this rate.


\subsection{Conclusion}

We ran multiple experiments to find what would be the best parameters for the
AF\_XDP sockets and the related application's datapath to lower the end-to-end
latency. We identified clusters of configurations giving similar latencies with
a 1 second inter batch interval and found that those combinations stayed
relevant at higher packet rates. We finally found minimal latencies of
\mus{6.5} for Mellanox and \mus{9.7} for Intel as well as the corresponding
configurations. All these measurements include a measurement overhead estimated
at $\sim$\mus{5}.

    \section{Related Work}
\label{sec:sota}

The impact of the kernel protocol stack implementation on the packet processing delay has been analyzed in details~\cite{caiUnderstandingHostNetwork2021}, and it has been shown that the usual packet processing pipeline can be slow, especially for packets needing several rounds of processing, such as packets in a VxLAN tunnel~\cite{zhuoSlimOSKernel2019}. To alleviate the problem, different strategies has been proposed, such as custom overlay networks aiming at bypassing the network stack~\cite{zhuoSlimOSKernel2019, CiliumWebsite}.

One widely accepted solution to the slow packet processing at the OS kernel is, with support of the Network Interface Cards, to take a packet at the earliest stage (basically, after reception of a packet by a NIC port) and send it immediately to the user space of a system. To do so, two different solutions exist today:
the Data Plane Development Kit and the eXpress Data Path.


To assess the benefits of DPDK and XDP, the networking community has carried out extensive experimental studies to compare the performance of DPDK vs the Linux kernel~\cite{karlssonPathDPDKSpeeds, tuRevisitingOpenVSwitch2021}, XDP vs the Linux kernel~\cite{ozturk, hoiland-jorgensenEXpressDataPath2018} and DPDK vs XDP~\cite{ozturk, hoiland-jorgensenEXpressDataPath2018}. Those papers report that XDP can achieve higher throughput than DPDK on a multi-core setting as well as providing more flexibility to integrate into an application on Linux.

A deep understanding of the packet processing path is important to chase out microseconds delays able to impact current latency sensitive cloud applications, frequently deployed in the form of chains of microservices \cite{barroso17, Gan19-deathstarbench}. Consequently, in this paper we go one step further and carry out an experimental study to understand the nature of latency added at every stage of the AF\_XDP architecture by exploring multiple parameters related to the socket, the NIC driver and some system parameters. As far as we know, this kind of study has not been reported yet in the literature.



    \section{Conclusions}
\label{sec:conc}

The eXpress Data Path and AF\_XDP are a promising solution to the slow packet processing problem in Linux's kernel space.
In this paper, we presented a comprehensive study on the performance of AF\_XDP from the latency point of view, as we believe, it is a key point to chase out microsecond delays and preserve the Quality of Service of latency sensitive applications.
Employing \texttt{xdpsock}, the standard AF\_XDP testing program, we conducted a set of experiments to analyze every optimization parameters of the AF\_XDP socket, which allows to easily exchange data packet between a user space program and the driver NIC.


We have run a configuration for every combination of parameters available with the AF\_XDP socket and the sample AF\_XDP application. We then identified different clusters of configurations leading to similar results and summarized what parameters lead to those different clusters and what where the latencies we obtained with them. We have also shown that we can achieve below \mus{10} latency using a basic AF\_XDP application when there is no buffering involved.

The results shown here are part of an ongoing project to understand the internals of AF\_XDP. As a next step, we plan to evaluate the impact of AF\_XDP on the energy consumption of servers, before and after the multiple optimizations, and evaluate the impact of the parameter optimization on real applications deployed as a chain of microservices, especially for data streaming.
    
    \section{Acknowledgements}

This work has been supported by the French government, through the EUR DS4H 
Investments in the Future project managed by the National Research Agency (ANR)
with the reference number ANR-17-EURE-0004.

    \bibliographystyle{IEEEtran}
    \bibliography{tex/biblio}

\begin{thebibliography}{10}
\providecommand{\url}[1]{#1}
\csname url@samestyle\endcsname
\providecommand{\newblock}{\relax}
\providecommand{\bibinfo}[2]{#2}
\providecommand{\BIBentrySTDinterwordspacing}{\spaceskip=0pt\relax}
\providecommand{\BIBentryALTinterwordstretchfactor}{4}
\providecommand{\BIBentryALTinterwordspacing}{\spaceskip=\fontdimen2\font plus
\BIBentryALTinterwordstretchfactor\fontdimen3\font minus
  \fontdimen4\font\relax}
\providecommand{\BIBforeignlanguage}[2]{{%
\expandafter\ifx\csname l@#1\endcsname\relax
\typeout{** WARNING: IEEEtran.bst: No hyphenation pattern has been}%
\typeout{** loaded for the language `#1'. Using the pattern for}%
\typeout{** the default language instead.}%
\else
\language=\csname l@#1\endcsname
\fi
#2}}
\providecommand{\BIBdecl}{\relax}
\BIBdecl

\bibitem{Gan19-deathstarbench}
\BIBentryALTinterwordspacing
Y.~Gan, Y.~Zhang, D.~Cheng, A.~Shetty, P.~Rathi, N.~Katarki, A.~Bruno, J.~Hu,
  B.~Ritchken, B.~Jackson, K.~Hu, M.~Pancholi, Y.~He, B.~Clancy, C.~Colen,
  F.~Wen, C.~Leung, S.~Wang, L.~Zaruvinsky, M.~Espinosa, R.~Lin, Z.~Liu,
  J.~Padilla, and C.~Delimitrou, ``An open-source benchmark suite for
  microservices and their hardware-software implications for cloud \& edge
  systems,'' in \emph{Proceedings of the Twenty-Fourth International Conference
  on Architectural Support for Programming Languages and Operating Systems},
  ser. ASPLOS '19.\hskip 1em plus 0.5em minus 0.4em\relax New York, NY, USA:
  Association for Computing Machinery, 2019, p. 3–18. [Online]. Available:
  \url{https://doi.org/10.1145/3297858.3304013}
\BIBentrySTDinterwordspacing

\bibitem{Li22-livenet}
\BIBentryALTinterwordspacing
J.~Li, Z.~Li, R.~Lu, K.~Xiao, S.~Li, J.~Chen, J.~Yang, C.~Zong, A.~Chen, Q.~Wu,
  C.~Sun, G.~Tyson, and H.~H. Liu, ``Livenet: A low-latency video transport
  network for large-scale live streaming,'' in \emph{Proceedings of the ACM
  SIGCOMM 2022 Conference}, ser. SIGCOMM '22.\hskip 1em plus 0.5em minus
  0.4em\relax New York, NY, USA: Association for Computing Machinery, 2022, p.
  812–825. [Online]. Available: \url{https://doi.org/10.1145/3544216.3544236}
\BIBentrySTDinterwordspacing

\bibitem{barroso17}
\BIBentryALTinterwordspacing
L.~Barroso, M.~Marty, D.~Patterson, and P.~Ranganathan, ``Attack of the killer
  microseconds,'' \emph{Commun. ACM}, vol.~60, no.~4, p. 48–54, mar 2017.
  [Online]. Available: \url{https://doi.org/10.1145/3015146}
\BIBentrySTDinterwordspacing

\bibitem{zhangDemikernelDatapathOS2021}
I.~Zhang, A.~Raybuck, P.~Patel, K.~Olynyk, J.~Nelson, O.~S.~N. Leija,
  A.~Martinez, J.~Liu, A.~K. Simpson, S.~Jayakar, P.~H. Penna, M.~Demoulin,
  P.~Choudhury, and A.~Badam, ``The {{Demikernel Datapath OS Architecture}} for
  {{Microsecond-scale Datacenter Systems}},'' in \emph{Proceedings of the {{ACM
  SIGOPS}} 28th {{Symposium}} on {{Operating Systems Principles}}}, ser.
  {{SOSP}} '21.\hskip 1em plus 0.5em minus 0.4em\relax {New York, NY, USA}:
  {Association for Computing Machinery}, Oct. 2021, pp. 195--211.

\bibitem{zhuoSlimOSKernel2019}
D.~Zhuo, K.~Zhang, Y.~Zhu, H.~H. Liu, M.~Rockett, A.~Krishnamurthy, and
  T.~Anderson, ``Slim: \{\vphantom\}{{OS}}\vphantom\{\} {{Kernel Support}} for
  a \{\vphantom\}{{Low-Overhead}}\vphantom\{\} {{Container Overlay Network}},''
  in \emph{16th {{USENIX Symposium}} on {{Networked Systems Design}} and
  {{Implementation}} ({{NSDI}} 19)}, 2019, p. 331.

\bibitem{DPDKWebsite}
``{{DPDK Website}},'' https://www.dpdk.org/.

\bibitem{XDPProjectGitHub}
``{{XDP Project GitHub}} profile,'' https://github.com/xdp-project.

\bibitem{vxlanRFC}
M.~Mahalingam, D.~Dutt, K.~Duda, P.~Agarwal, L.~Kreeger, T.~Sridhar,
  M.~Bursell, and C.~Wright, ``Virtual {{eXtensible Local Area Network}}
  ({{VXLAN}}): {{A Framework}} for {{Overlaying Virtualized Layer}} 2
  {{Networks}} over {{Layer}} 3 {{Networks}},'' {Internet Engineering Task
  Force}, Request for {{Comments}} RFC 7348, Aug. 2014.

\bibitem{tuRevisitingOpenVSwitch2021}
W.~Tu, Y.-H. Wei, G.~Antichi, and B.~Pfaff, ``Revisiting the open {{vSwitch}}
  dataplane ten years later,'' in \emph{Proceedings of the 2021 {{ACM SIGCOMM}}
  2021 {{Conference}}}, ser. {{SIGCOMM}} '21.\hskip 1em plus 0.5em minus
  0.4em\relax {New York, NY, USA}: {Association for Computing Machinery}, Aug.
  2021, pp. 245--257.

\bibitem{FacebookincubatorKatran2023}
``{{GitHub}} repository of {{Katran}} by {{Meta}},'' Meta Incubator, Feb. 2023.

\bibitem{karlssonPathDPDKSpeeds}
M.~Karlsson and B.~Topel, ``The {{Path}} to {{DPDK Speeds}} for {{AF XDP}}.''

\bibitem{AFXDPLinux}
``{{AF}}\_{{XDP}} \textemdash{} {{The Linux Kernel}} documentation,''
  https://www.kernel.org/doc/html/latest/networking/af\_xdp.html.

\bibitem{NAPIPaper}
\BIBentryALTinterwordspacing
J.~H. Salim, R.~Olsson, and A.~Kuznetsov, ``Beyond softnet,'' in \emph{5th
  Annual Linux Showcase \& Conference (ALS 01)}.\hskip 1em plus 0.5em minus
  0.4em\relax Oakland, CA: USENIX Association, Nov. 2001. [Online]. Available:
  \url{https://www.usenix.org/conference/als-01/beyond-softnet}
\BIBentrySTDinterwordspacing

\bibitem{NAPI}
``{NAPI} article on the {Linux} {Foundation's} wiki,''
  https://wiki.linuxfoundation.org/networking/napi.

\bibitem{perfsCoalescing}
K.~Salah, K.~{El-Badawi}, and F.~Haidari, ``Performance analysis and comparison
  of interrupt-handling schemes in gigabit networks,'' \emph{Computer
  Communications}, vol.~30, no.~17, pp. 3425--3441, Nov. 2007.

\bibitem{AFXDPZC}
``{{AF}}\_{{XDP}}: Introducing zero-copy support [{{LWN}}.net],''
  https://lwn.net/Articles/756549/.

\bibitem{afxdpBPRFC}
M.~Karlsson, ``{RFC} for busy poll support for {AF\_XDP} sockets,''
  \url{https://lore.kernel.org/bpf/CAJ8uoz3MNrz\_f7dy6+U=zj7GeywUda9E9pP2s9uU1jVW5O2zHw@mail.gmail.com/T/}.

\bibitem{xdpsockGithub}
``xdpsock.c github link,''
  \url{https://github.com/xdp-project/bpf-examples/blob/eeb154d7a1c6d155e7d8d9da9018ceeff3ebfaf4/AF_XDP-example/xdpsock.c}.

\bibitem{UsingUserspaceTracepoints}
``{Using} {User-Space} {Tracepoints} with {{BPF}} [{{LWN}}.net],''
  https://lwn.net/Articles/753601/.

\bibitem{PerfWiki}
``Perf {{Wiki}},'' https://perf.wiki.kernel.org/index.php/Main\_Page.

\bibitem{hoiland-jorgensenEXpressDataPath2018}
T.~{H{\o}iland-J{\o}rgensen}, J.~D. Brouer, D.~Borkmann, J.~Fastabend,
  T.~Herbert, D.~Ahern, and D.~Miller, ``The {{eXpress}} data path: Fast
  programmable packet processing in the operating system kernel,'' in
  \emph{Proceedings of the 14th {{International Conference}} on Emerging
  {{Networking EXperiments}} and {{Technologies}}}, ser. {{CoNEXT}} '18.\hskip
  1em plus 0.5em minus 0.4em\relax {New York, NY, USA}: {Association for
  Computing Machinery}, Dec. 2018, pp. 54--66.

\bibitem{rizzoNetmapNovelFramework2012}
L.~Rizzo, ``Netmap: {{A Novel Framework}} for {{Fast Packet}}
  \{\vphantom\}{{I}}/{{O}}\vphantom\{\},'' in \emph{2012 {{USENIX Annual
  Technical Conference}} ({{USENIX ATC}} 12)}, 2012, pp. 101--112.

\bibitem{Ansible}
``Ansible official website,'' https://www.ansible.com.

\bibitem{KDE}
``Seaborn's documentation on {{KDE}},''
  https://seaborn.pydata.org/generated/seaborn.kdeplot.html.

\bibitem{caiUnderstandingHostNetwork2021}
Q.~Cai, S.~Chaudhary, M.~Vuppalapati, J.~Hwang, and R.~Agarwal, ``Understanding
  host network stack overheads,'' in \emph{Proceedings of the 2021 {{ACM
  SIGCOMM}} 2021 {{Conference}}}, ser. {{SIGCOMM}} '21.\hskip 1em plus 0.5em
  minus 0.4em\relax {New York, NY, USA}: {Association for Computing Machinery},
  Aug. 2021, pp. 65--77.

\bibitem{CiliumWebsite}
``Cilium - {{Linux Native}}, {{API-Aware Networking}} and {{Security}} for
  {{Containers}},'' https://cilium.io.

\bibitem{ozturk}
T.~{\"O}zturk, ``Performance {{Evaluation}} of {{eXpress Data Path}} for
  {{Container-Based Network Functions}},'' p.~65.

\end{thebibliography}
\end{document}